\documentclass[a4paper,11pt]{article}

\usepackage[left=2.5cm,right=2.5cm,top=2.5cm,bottom=2.5cm]{geometry}
\usepackage{graphicx,amssymb,amsmath,amsthm,mathrsfs,setspace,subcaption,cite,authblk,float,multicol}

\usepackage[colorlinks=true,bookmarks=true]{hyperref}
\usepackage{orcidlink}

\usepackage{mathtools}

\DeclarePairedDelimiterX\braket[2]{\langle}{\rangle}{#1 \delimsize\vert #2}

\hypersetup{citecolor=blue}
\newcommand{\dif}{\mathrm{d}}
\newcommand{\Eqref}[1]{(\ref{#1})}
\newcommand{\half}{\frac{1}{2}}

\newcommand{\brac}[1]{\left(#1 \right)}
\newcommand{\sbrac}[1]{\left[#1\right]}

\newenvironment{remark}[1][Remark]{\begin{trivlist}
\item[\hskip \labelsep {\bfseries #1}]}{\end{trivlist}}

\numberwithin{equation}{section}
 
\begin{document}

\title{Interior marginally outer trapped surfaces in Hayward black holes}

\author[1]{Abbas M.~Sherif\footnote{Email: abbasmsherif25@gmail.com}\orcidlink{0000-0002-4038-6628}}

\author[2]{Yen-Kheng Lim\footnote{Email: yenkheng.lim@gmail.com, yenkheng.lim@xmu.edu.my}\orcidlink{0000-0002-0907-6904}}

\affil[1]{\normalsize{\textit{Institute of Mathematics, Henan Academy of Sciences (HNAS), 228 Mingli Road, Zhengzhou 450046, Henan, China}}}

\affil[2]{\normalsize{\textit{Department of Physics, Xiamen University Malaysia, Jalan Sunsuria, 43900 Sepang, Malaysia}}}

\date{\normalsize{\today}}

\renewcommand\Authands{, and }

\maketitle

\begin{abstract}
We locate interior marginally outer trapped and marginally outer trapped open surfaces (MOTS/MOTOS) in the regular Haward metric with parameter $b$ for which a critical (extremal) value $b=b_c$ demarcates when the spacetime admits no horizon and when it admits inner and outer horizons. We identify self-intersecting MOTS which occur in pairs. For $b$ close to the critical value, there are no self-intersecting MOTS/MOTOS, and one can fine-tune $b$ so that the interior contains only near-spherical MOTS. We also show that in a neighborhood of the inner horizon for certain values of $b$, upon reduction of the problem to a singular Sturm-Liouville problem, the locations of the MOTS are given by hypergeometric functions,  the eigenspace of the operator for which is complete, not discrete, and discontinuous.
\end{abstract}


\section{Introduction} \label{sec_intro}


Black holes provide a fertile ground for merging concepts of General Relativity and Quantum Mechanics. It is therefore not at all surprising that the study of black holes has permeated various areas of research in theoretical and mathematical physics. They are characterized by trapped surfaces, closed 2-surfaces with convergent orthogonally emitted light rays. These surfaces are enclosed by a boundary 3-manifold called a ``horizon'', and their presence signals the inevitable formation of singularities in a spacetime under physically reasonable conditions \cite{Penrose:1969}.

Originally, much of what was understood about black holes came from information related to the event horizon of static (and by extension, stationary) black holes \cite{Hawking:1973}. The global description of these objects implied that the evolution of any black hole enclosed by an event horizon cannot be observed. This problem led to the introduction of local notions of horizons in terms of marginally outer trapped surfaces (MOTS), the ``marginal'' case of a trapped surface. On these surfaces outgoing light rays neither converge nor diverge. The local character of these objects has led to their widespread applications, even in highly dynamical situations, for describing black holes \cite{Ashtekar:2004cn,Ashtekar:2005some,booth:2005black,hayward:1994general,booth:2004first,sherif:2019some} (also see the recent work \cite{Ashtekar:2025wnu} for a comprehensive review on the subject). They have been used to recover the laws of black hole mechanics for dynamical spacetimes.

While any hypersurface enclosing a black hole has the property of a MOTS, a surface having the property of a MOTS need not enclose any black hole. Trivial examples are the inner horizons of spacetimes admitting a pair of horizons. In such a case trapped surfaces are located to the ``outside'' of the inner horizon rather than to its ``inside''. The reason is that enclosing a black hole requires additional constraints on the expansion of the null geodesics that point into the MOTS, as well as the variation of the outward expansion along the inward pointing null geodesics. It is then only natural to expect an abundance of surfaces satisfying just the MOTS property.

The MOTS equation whose solutions generate these more general surfaces have been termed \emph{MOTSodesic equations} and the solutions are \emph{MOTSodesic curves}, mimicking the geodesic equations, and the MOTS is obtained as a surface of revolution of the MOTSodesic curves about a symmetry axis. Some of these surfaces exhibit interesting topological properties and self-intersections. They have been found in a variety of static and stationary spacetimes \cite{Booth:2020qhb,Booth:2021sow,Sievers:2023zng,Booth:2022vwo} and have been shown to play crucial roles in the mergers of black holes \cite{pook:2021happens,pook:2021ultimate,Booth:2021sow}. MOTSodesic curves generate MOTS when closed and MOT(Open)S when open.

The evolution of a MOTS to a horizon enclosing a black hole may be determined by a notion of stability  introduced by Andersson {\em et al.} \cite{Andersson:2005gq,Andersson:2007fh} as a property of an elliptic operator. When the principal eigenvalue, that with the smallest real part, is non-negative or strictly positive, the MOTS is said to be stable or strictly stable, but otherwise unstable. The main results of \cite{Andersson:2005gq} essentially posits that if the eigenvalue spectrum is positive, the MOTS smoothly evolves to a hypersurface, called a marginally outer trapped tube (MOTT), and will enclose trapped surfaces if, in addition, certain curvature conditions are met.

Generically, interior MOTS/MOTOS are unstable \cite{pook:2021happens,pook:2021ultimate,Booth:2021sow,Booth:2022vwo,Booth:2023aan}, and because of the prevalence of these MOTS and the identification of their growing relevance, interests in understanding the nature of their stability is also growing \cite{Booth:2023aan,Hennigar:2021ogw,bussey:2021eigenvalues,sherif:2025instability}. In fact, one important result of \cite{Booth:2023aan} is the association of the number of self-intersections of a MOTS with a lower bound on the number of negative eigenvalues of the stability operator. For a region where there are no self-intersecting MOTS, the MOTS can be seen as deformed spheres. Where these MOTS are in the near-horizon region, one may obtain them by perturbing the MOTSodesic equations about a spherical horizon and obtaining analytic solutions in terms of known special (an in some cases elementary) functions. As will be seen for the metric of interest, the common practice of employing Legendre functions does not always work as the solutions to the perturbed system are constrained by the particular nature the metric parametrization.

In this paper, we study the properties of MOTS and MOTOS in the Hayward black hole. The Hayward metric is a regular non-vacuum solution to the Einstein equations with a parameter that achieves a critical value below which there is a pair of horizons -- an inner and an outer one -- above which there is no black hole, and at which there is a single extremal horizon. We aim to examine the nature of interior MOTS in the Hayward metric as one varies the Hayward parameter. In particular, how does the number of self-intersecting MOTS, as well as the number of self-intersections of a MOTS depend on the value of the parameter relative to that value where the parameter is critical? This relationship provides important insights into the character of the MOTSodesic equations.

This paper is structured as follows. In Sec.~\ref{Sec:PG} we introduce the notion of a marginally outer trapped surface (MOTS) and the Painlev\'{e}-Gullstrand (PG) spacetime slicing which we use to locate MOTS to the interior of a black hole horizon. We then describe, in an abridged way, how to construct axi-symmetric surfaces in a constant PG time slice and derive the associated MOTS equation (a more complete description is provided in Appendix A). In Sec.~\ref{Solutions}, we describe how to numerically generate solutions to the MOTSodesic equations for the Hayward metric. The impact of the Hayward parameter on the behavior of the solution curves is comprehensively discussed. In a near-horizon neighborhood we obtain analytic solutions to the MOTSodesic equations and demonstrate its agreement with the numerical solutions. We conclude in Sec.~\ref{Conclusion} with a discussion of our results.


\section{MOTS in Painlev\'{e}-Gullstrand slices}\label{Sec:PG}


Consider a spacetime $\mathcal{M}$ with Lorentzian metric $g_{\mu\nu}$ that is foliated by spacelike slices $\{\Sigma_t\}$ with unit normal $u^{\mu}$. Given a closed 2-surface $\mathcal{S}$ with unit normal $n^{\mu}$ in a slice $\Sigma$ of $M$, one defines null vector fields
\begin{align}
k^{\mu}=u^{\mu}+n^{\mu},\quad l^{\mu}=u^{\mu}-n^{\mu},\label{nullgauge}
\end{align}
spanning $T^{\perp}(\mathcal{S})$, so that the induced metric on $\mathcal{S}$ is $q_{\mu\nu}=g_{\mu\nu}+k_{(\mu}l_{\nu)}$, with the usual symmetrization notation. We orient $\mathcal{S}$ such that $k^{\mu}$ is taken to point outward. The surface $\mathcal{S}$ is a marginally outer trapped surface (MOTS) if the 2-divergence of $k^{\mu}$, $\theta_k$, called the \emph{outward null expansion} and given by the sum of the mean curvature of $\mathcal{S}$ and the extrinsic curvature of $S$ along $u^a$, vanishes everywhere on $\mathcal{S}$ \cite{hayward:1994general,Ashtekar:2004cn,Ashtekar:2005some,booth:2005black}. There is a gauge freedom to boost the null pair, $\{k^{\mu}\rightarrow\tilde f k^{\mu},l^{\mu}\rightarrow\tilde f^{-1}l^{\mu}\}$, with $\tilde f$ taken to be positive to preserve the orientation, for which the MOTS condition is invariant.

The vanishing condition $\theta_k=0$ is now known to be satisfied by an infinite set of surfaces in the black hole interior as referenced earlier. Switching coordinates allows one to probe this property to the interior of the black hole. The compactness property associated with the vanishing condition is dropped, and only when a surface satisfies closure that it is termed a MOTS. As such the MOTS condition also gives rise to open surfaces called MOTOS \cite{Booth:2020qhb}. Especially to examine surfaces satisfying the vanishing condition in the interior of a black hole, it is useful to switch to the Painlev\'{e}-Gullstrand (PG) coordinate system \cite{Booth:2020qhb}.

For any static spherically symmetric metric of the coordinate form
\begin{align}
 \dif s^2=-f(r)\dif t^2+f^{-1}\dif r^2+r^2\dif\Omega^2,\label{static1}
\end{align}
with $f=f(r)$ and $\dif\Omega^2=\dif\theta^2+\sin^2\theta\,\dif\phi^2$ being the canonical unit sphere metric, the change of coordinate $\tau=t+B(r)$, with the choice of function $B$ such that $B'=-f^{-1}\sqrt{1-Cf}$, for some arbitrary function $C=C(r)>0$, brings the metric of \Eqref{static1} to the PG-type form
\begin{align}
 \dif s^2=-f\dif\tau^2+2\sqrt{1-Cf}\dif\tau\dif r+C\dif r^2+r^2\dif\Omega^2,\label{PGstatic1}
\end{align}
where the prime denotes differentiation with respect to the coordinate $r$. Aspects of the case of a general $C$ has been addressed previously in \cite{Hennigar:2021ogw}, and in the case of a constant $C$, this requires $f<1$. The constant $\tau$ hypersurfaces are intrinsically flat, but with non-zero extrinsic curvature, allowing the use of some standard approaches in Euclidean flat space.

To construct an axi-symmetric MOTS $\mathcal{S}$ in a PG $\tau$-slice $(\Sigma,h_{ab},K_{ab},D_a)$ as a surface of revolution of a curve $\gamma:\mathbb{R}\longrightarrow\Sigma$ in $\Sigma$, with $K_{ab}$ and $D_a$ being the extrinsic curvature and compatible covariant derivative, let us suppose that $\Sigma$ admits a $U(1)$ symmetry generated by a vector field $\zeta^a$. Introduce an arclength parametrization of the curve $\gamma(\lambda)$ with tangent vector $v^a$ which we impose to have unit norm (this $v_av^a=1$ condition can be seen as a gauge choice as it is compatible with the normalizations of the null vectors). One generates the outward pointing unit normal $n^a$ to $\mathcal{S}$ in $\Sigma$ upon revolution about the symmetry axis by taking the cross product of $\zeta^a$ and $v^a$. One finds that the extrinsic curvature of $S$ along $u^a$ and the mean curvature of $\mathcal{S}$ are respectively given by
\begin{align}
z_1&=q^{ab}K_{ab},\label{extrinsiccurvatureS}\\
z_2&=-\kappa+n^a\partial_a\ln R,\label{meancurvatureS}
\end{align}
where $\kappa=n^av^b\mathcal{D}_bv_a$ is the magnitude of the acceleration of $v^a$, $\mathcal{D}_a$ is the compatible derivative associated to the induced metric $q_{ab}$ on $\mathcal{S}$, and $R$ here is denoting the component of $\zeta^a$, so that the MOTS condition is now given by the expression
\begin{align}
\kappa=n^a\partial_a\ln R+q^{ab}K_{ab}.\label{motscondition1}
\end{align}

Explicitly, the MOTS condition is, for the metric \Eqref{PGstatic1} with $C=1$, given by the following pair of coupled ordinary differential equations
\begin{subequations} \label{ddotr_ddottheta}
\begin{align}
 \ddot{r}&=r\dot{\theta}^2+\kappa r\dot{\theta},\label{EOM_rddot_c1}\\
 \ddot{\theta}&=-\frac{2}{r}\dot{r}\dot{\theta}-\kappa\frac{\dot{r}}{r},\label{EOM_thetaddot_c1}\\
\kappa&=\frac{r\dot{r}^2f'-2(1-f)(1+r^2\dot{\theta}^2)}{2r\sqrt{(1-f)}}\nonumber\\
&-\frac{1}{r}\brac{\dot{r}\cot\theta-r\dot{\theta}}.\label{kappa_eqn_c1}
\end{align}
\end{subequations}
(See Appendix A for a detailed discussion on the construction of these axisymmetric MOTS.) The overhead dot denotes differentiation with respect to the curve parameter $\lambda$. These equations are referred to as the \emph{MOTSodesic equations}, with solution curves being the \emph{MOTSodesics} in analogy with geodesics.


\section{MOTS/MOTOS in Hayward metric}\label{Solutions}


The Hayward metric is a regular metric given by Eq.~\Eqref{static1} with
\begin{align}
 f(r)&=1-\frac{2mr^2}{r^3+2mb^2}.\label{hayward1}
\end{align}
This is a solution to Einstein's equation $R_{\mu\nu}-\half R g_{\mu\nu}=8\pi T_{\mu\nu}$ where $T_{\mu\nu}=\brac{\rho+p_r}U_\mu U_\nu+p_\perp g_{\mu\nu}+(p_r-p_\perp)N_\mu N_\nu$ is the stress tensor of a co-moving anisotropic fluid of four-velocity $U_\mu=\brac{-\sqrt{f},0,0,0}$ and direction of anisotropy $N_\mu=\brac{0,1/\sqrt{f},0,0}$. Explicitly, the mass density $\rho=-{T^t}_t$, radial pressure $p_r={T^r}_r$, and tangential pressures $p_\perp={T^\theta}_\theta={T^\phi}_\phi$ are given by
\begin{align*}
\rho=-p_r&=\frac{3m^2b^2}{2\pi \brac{r^3+2mb^2}^2},\\
p_\perp&=\frac{3m^2b^2\brac{r^3-mb^2}}{\pi\brac{r^3+2mb^2}^3}.
\end{align*}
The Hayward spacetime has two distinct horizons for $b<(4/3\sqrt{3})\simeq0.7698m$, a degenerate horizon for $b=(4/3\sqrt{3})m$, and is horizonless for $b>(4/3\sqrt{3})m$. $b=0$ recovers the Schwarzschild solution.

\subsection{Numerical solutions}

We will now numerically solve the MOTSodesic equation for the Hayward regular black hole, examining the effects of the parameter $b$ on the nature of the MOTSodesics. The MOTSodesics equations are numerically solved using the fourth-order Runge--Kutta procedure implementd in C.

To find self-intersecting MOTSodesics, we wish to start the evolution of the MOTSodesic equations at the $z$-axis. Our initial conditions are therefore
\begin{align}
 r(0)=r_0,\quad\theta(0)=0,\quad\dot{r}(0)=0, \label{Hayward_init_cond}
\end{align}
for some choice of $r_0$ and $\dot{\theta}(0)$ being determined from the previous three conditions via the normalization condition Eq.~\Eqref{first_integral}. We then apply a shooting method to find curves that terminate at the $z$-axis at some final parameter $\lambda_{\mathrm{final}}$ such that $\theta(\lambda_{\mathrm{final}})=\pi$, which will ensure a compact surface upon revolving about the $z$-axis. If furthermore we have $\dot{r}(\lambda_{\mathrm{final}})=0$, the surface is regular.

\begin{figure}[ht!]
 \centering
 \includegraphics[scale=1]{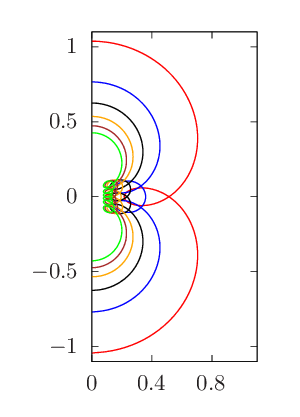}
 \caption{The first six self-intersecting MOTS in the Schwarzschild ($b=0$) case.}
 \label{fig_SchMOTS}
\end{figure}

From the second term of Eq.~\Eqref{kappa_eqn} the numerics is problematic when evaluated directly at $\lambda=0$ due to the $\sin\theta$ term in the denominator in $\kappa$. To circumvent this, we perform an expansion about \Eqref{Hayward_init_cond} by writing, up to second order in $\lambda$,
\begin{align}
 r(\lambda)=r_0+a\lambda^2,\quad\theta(\lambda)=\alpha\lambda+\beta\lambda^2, \label{Hayward_pole_exp}
\end{align}
for some constants $a$, $\alpha$, and $\beta$. In this form, the $\lambda\rightarrow0$ limit is consistent with the initial conditions \Eqref{Hayward_init_cond}. Substituting Eq.~\Eqref{Hayward_pole_exp} into \Eqref{ddotr_ddottheta}, we find
\begin{align}
 \beta&=0,\quad a=r_0\alpha\mathfrak{g},\nonumber\\
\mathfrak{g}&=\frac{\sbrac{2\alpha\sqrt{m(r_0^3+2mb^2)}-\sqrt{2}m\brac{1+r_0^2\alpha^2}}}{4\sqrt{m(r_0^3+2mb^2)}}.
\end{align}
The constraint equation, up to second order in $\lambda$, is
\begin{align}
 r_0^2\alpha^2+\brac{4a^2+2r_0a\alpha^2}\lambda^2\simeq1. \label{first_integral_exp}
\end{align}
Hence, since the numerics are problematic at $\lambda=0$, we choose a small $\lambda=\epsilon$ and determine $\alpha$ using Eq.~\Eqref{first_integral_exp}. With $\beta=0$, and $\alpha$ and $a$ determined for a choice of $m$ and $b$, we start our numerical evolution at
\begin{align}
 r(\epsilon)=r_0+\epsilon^2,\quad \theta(\epsilon)=\alpha\epsilon,\quad 0<\epsilon\ll 1. \label{numerical_init_cond}
\end{align}
In practice, one may choose $\lambda$ to be a non-zero number sufficiently small such that $\epsilon^2$ is below numerical precision, we have $\alpha\simeq 1/r_0$ and the starting position for numerical evolution is, to good approximation, $r(\epsilon)\simeq r_0$, $\theta(\epsilon)\simeq\epsilon/r_0$, and $\dot{\theta}(\epsilon)\simeq1/r_0$, where a typical choice of $\epsilon$ would be $\epsilon=10^{-5}$. As a test of our equations of motion and numerical solver, we verify that the Schwarzschild case $b=0$ reproduces the self-intersecting MOTS of \cite{Booth:2020qhb}. Several of these self-intersecting MOTS are shown in Fig.~\ref{fig_SchMOTS}.

\begin{figure}[ht!]
\centering
    \includegraphics[scale=0.8]{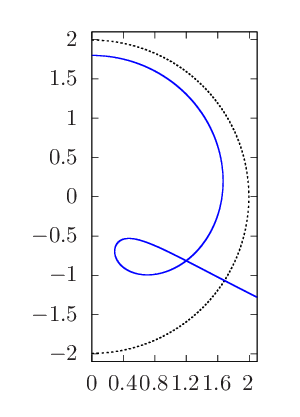}
    \includegraphics[scale=0.8]{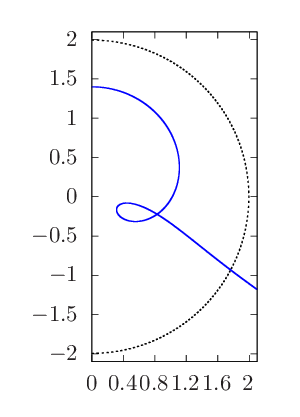}
    \includegraphics[scale=0.8]{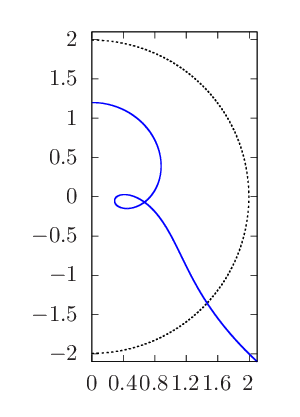}
\caption{$r_0=1.8m,1.4m,1.2m$, respectively.}
\label{fig_data_b0.10_rstart1.8}
\end{figure}

\begin{figure}[ht!]
\centering
    \includegraphics[scale=0.8]{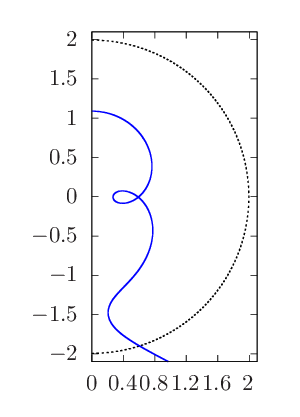}
    \includegraphics[scale=0.8]{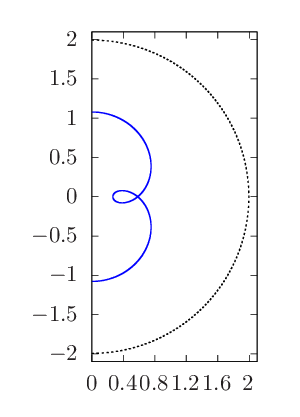}
    \includegraphics[scale=0.8]{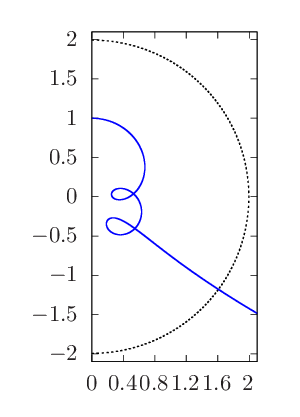}
\caption{$r_0=1.09m,1.0794m,1.0m$, respectively.}
\label{fig_data_b0.10_rstart1.09}
\end{figure}

\begin{figure}[ht!]
\centering
    \includegraphics[scale=0.8]{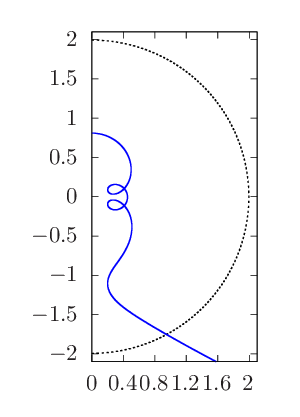}
    \includegraphics[scale=0.8]{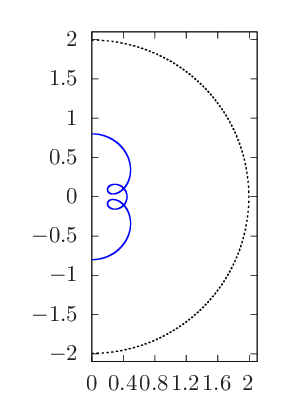}
    \includegraphics[scale=0.8]{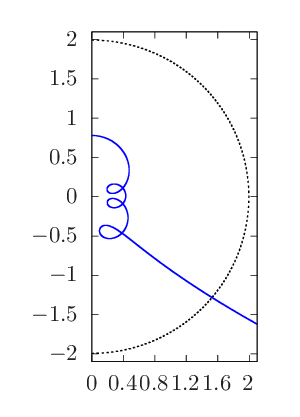}
\caption{$r_0=0.81m,0.80125m,0.78m$.}
\label{fig_data_b0.10_rstart0.81}
\end{figure}

In the following we explain the types of MOTS and MOTOS obtained from solving the MOTSodesic equations with the initial conditions \Eqref{Hayward_init_cond}, for varying choices of $r_0$. Let us proceed by starting from the outer horizon. Starting with $r_0$ coinciding with the outer horizon, we have the spherical MOTS as discussed in the previous section. Figs.~\ref{fig_data_b0.10_rstart1.8} to Fig.~\ref{fig_data_b0.10_rstart0.81} show some outer MOTS/MOTOS of the Hayward spacetime with $b=0.1m$, with different initial positions $r_0$ on the $z$-axis. In each panel the horizons are depicted as the black dotted semi-circles in the case $b=0.1m$, for which the outer horizon is located at $r\simeq1.9950m$. For values of $r_0$ just inside the outer horizon, we have MOTOS that loops and intersects itself before extending into infinity, shown in Fig.~\ref{fig_data_b0.10_rstart1.8} and the first panel of Fig.~\ref{fig_data_b0.10_rstart1.09}. For a specific value of $r_0$, the MOTSOdesic curve intersects the $z$-axis again at $\dot{r}=0$, thus forming a closed curve, i.e. a MOTS. This is shown in the second panel of Fig.~\ref{fig_data_b0.10_rstart1.09}. This is a MOTS with ``one loop''. Further decreasing $r_0$ has the MOTSodesic curve forming a second loop, which in general are MOTOS which extend to infinity. This is shown in the last panel of Fig.~\ref{fig_data_b0.10_rstart1.09} and the first panel of Fig.~\ref{fig_data_b0.10_rstart0.81}. A further decrease in $r_0$ again yields a MOTS of two loops, and then a MOTOS of three loops as seen respectively in the last two panels of Fig.~\ref{fig_data_b0.10_rstart0.81}.

\begin{figure}[ht!]
\centering
    \includegraphics[scale=0.8]{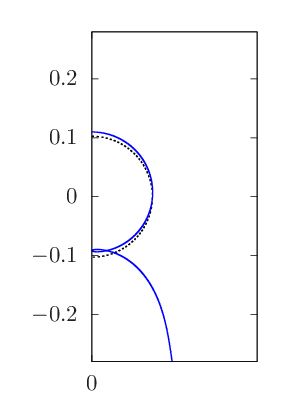}
    \includegraphics[scale=0.8]{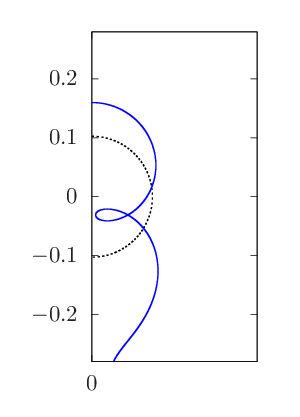}
    \includegraphics[scale=0.8]{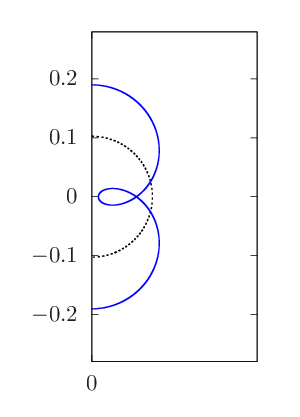}
\caption{$r_0=0.11m,0.16m,0.19m$, respectively.}
\label{fig_data_b0.10_rstart0.11}
\end{figure}

\begin{figure}[ht!]
\centering
    \includegraphics[scale=0.8]{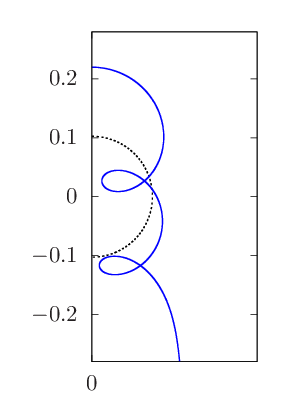}
    \includegraphics[scale=0.8]{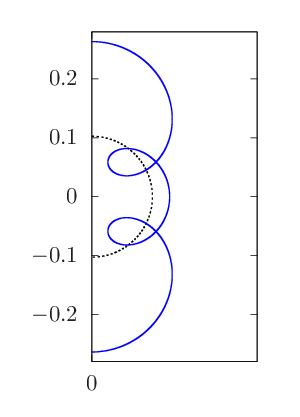}
    \includegraphics[scale=0.8]{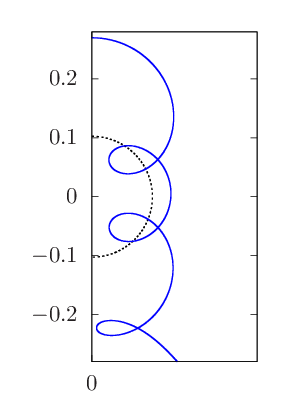}
\caption{$r_0=0.22m,0.2635m,0.26m$, respectively.}
\label{fig_data_b0.10_rstart0.22}
\end{figure}

Next, we explore the solutions starting from the inner horizon. Fig.~\ref{fig_data_b0.10_rstart0.11} and Fig.~\ref{fig_data_b0.10_rstart0.22} show some inner MOTS/MOTOS of the Hayward spacetime for the same value $b=0.1m$, with different initial positions $r_0$ on the $z$-axis. As expected, for $r_0$ coinciding with the inner horizon we have a spherical MOTS, shown as the black dotted semi-circles in each panel. Now, we increase $r_0$ to move away from the inner horizon. There we see the MOTSOdesic curve forming a single loop, shown in the first two panels of Fig.~\ref{fig_data_b0.10_rstart0.11}. In general, these are MOTOS which extend to infinity until $r_0$ reaches a particular value for which we get a `one-loop' closed MOTS, shown in the third panel of Fig.~\ref{fig_data_b0.10_rstart0.11}. Further increasing $r_0$ now has the curves forming a second loop. Again, these are generically MOTOS until $r_0$ reaches a value giving a ``two-loop'' MOTS, shown in the second panel of Fig.~\ref{fig_data_b0.10_rstart0.22}.

\begin{figure}[ht!]
\centering
    \includegraphics[scale=1]{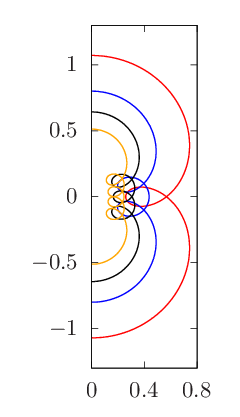}
    \includegraphics[scale=1]{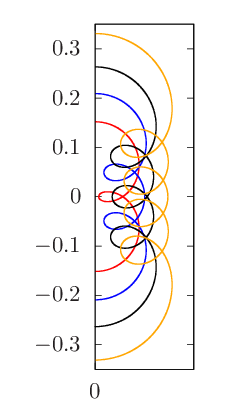}
\caption{Self-intersecting MOTS in the Hayward spacetime of $b=0.08m$. Each $n$-loop self-intersecting MOTS (here $1\leq n\leq 4$) occur in pairs, one close to the outer horizon (first panel) and another close to the inner horizon (second panel). Note the different axes scales of the two panels.}
  \label{fig_MOTSb0.08}
\end{figure}

In the preceding discussions, note that the closed MOTS occur in pairs; one associated to the outer horizon and the other associated to the inner horizon. This is similar to that of the Gauss--Bonnet black holes, also with two horizons \cite{Hennigar:2021ogw}. In particular, a MOTS with $n$ self-intersections occur in pairs, one associated with the outer horizon and the other associated with the inner horizon. For fixed $b$, the number of existing pairs depend on the particular value of $b$. As $b$ is increased, the inner horizon approaches the outer horizon and the number  of existing self-intersecting MOTS pairs decrease. Intuitively, one way to explain this is that the ``available space'' between the two horizons become ``smaller''. For instance, when $b=0.08m$ there are four pairs of self-intersecting MOTS with $1\leq n\leq 4$ self intersections,  shown in Fig.~\ref{fig_MOTSb0.08}.

\begin{figure}[ht!]
\centering
    \includegraphics[scale=0.8]{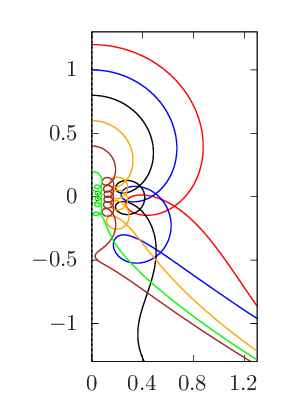}
    \includegraphics[scale=0.8]{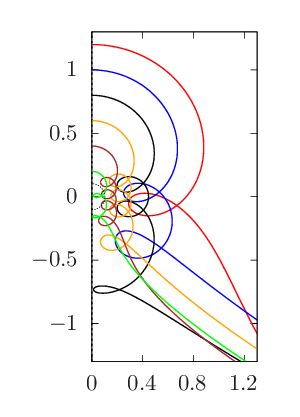}
    \includegraphics[scale=0.8]{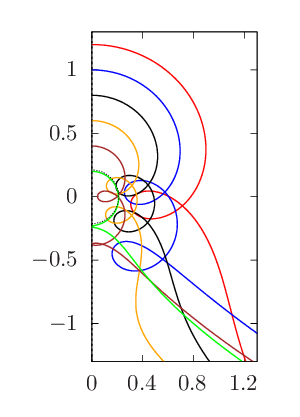}
\caption{$b=0.05m,0.1m,0.2m$, respectively.}
\label{fig_data_bstart0.05}
\end{figure}

\begin{figure}[ht!]
\centering
    \includegraphics[scale=0.8]{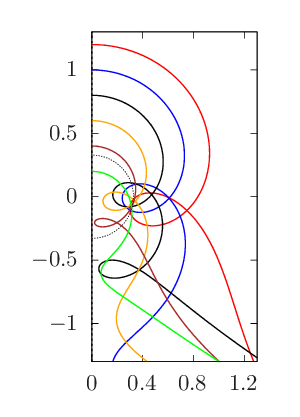}
    \includegraphics[scale=0.8]{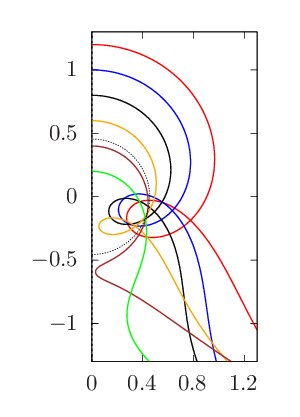}
    \includegraphics[scale=0.8]{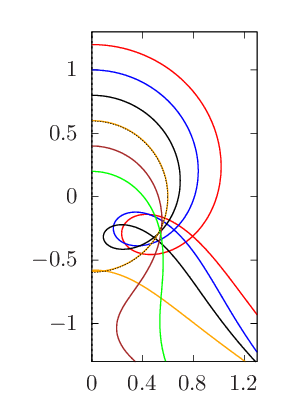}
\caption{$b=0.3m,0.4m,0.5m$, respectively.}
\label{fig_data_bstart0.3}
\end{figure}

We now plot in Figs.~\ref{fig_data_bstart0.05} and \ref{fig_data_bstart0.3}, for several $b$'s, MOTOS for starting positions $r_0$ equal to $1.2m$, $1.0m$, $0.8m$, $0.6m$, $0.4m$, and $0.2m$ for various values of $b$. against the background of the inner horizon. Due to space constraints, we include MOTOS for several initial values on a single plot for each $b$. Cluttered, but with clear enough visuals.

\subsection{Perturbing the MOTSodesic equations}

MOTS stability, via an elliptic operator and its spectrum, plays a crucial role in understanding the dynamical evolution of a black hole \cite{Andersson:2005gq,Andersson:2007fh}. The operator characterizes the deformation of a MOTS $\mathcal{S}$ along its normal direction in the embeddding slice. 
\begin{align}
L_{\mathcal{S}}&=-\mathcal{D}^2+\frac{1}{2}\left(\mathcal{R}-|\sigma|^2-2\mathcal{G}_{ab}k^au^b\right),\label{Stability_operator_1}
\end{align}
where the scalar curvature of the MOTS $\mathcal{R}$, the term involving the Einstein tensor $\mathcal{G}_{ab}$, and the norm of the shear $\sigma$ of $k^a$, in the case of the Hayward metric, are
\begin{align*}
\mathcal{R}&=2\kappa\brac{\frac{\dot{r}}{r}\cot\theta-\dot{\theta}},\quad\mathcal{G}_{ab}k^au^b=\Psi\brac{\dot{r}^2-1},\\
\Psi&=-\frac{12m^2b^2}{(r^3+2mb^2)^2},\quad|\sigma|^2=\mathcal{Q}^2+\frac{1}{r^2}\dot{r}^2\cot^2\theta,\\
\mathcal{Q}&=2\dot{r}^2\dot{\theta}+\frac{f'\dot{r}^2}{2\sqrt{1-f}}-r\sqrt{1-f}\;\dot{\theta}^2.
\end{align*}
Solving the eigenvalue problem
\begin{align}
L_{\mathcal{S}}\psi=\tilde\lambda\psi,\label{st3}
\end{align}
for the eigenvalue $\tilde\lambda_0$ with the smallest real part,for some positive function $\psi$, characterizes stability: $\tilde\lambda_0\geq0$ implies stability and $\tilde\lambda_0>0$ implies strict stability. Otherwise, $\mathcal{S}$ is unstable.

\begin{figure}[ht!]
\centering
    \includegraphics[scale=0.9]{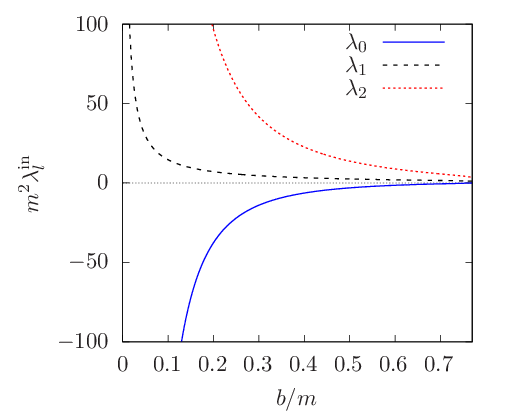}
    \includegraphics[scale=0.9]{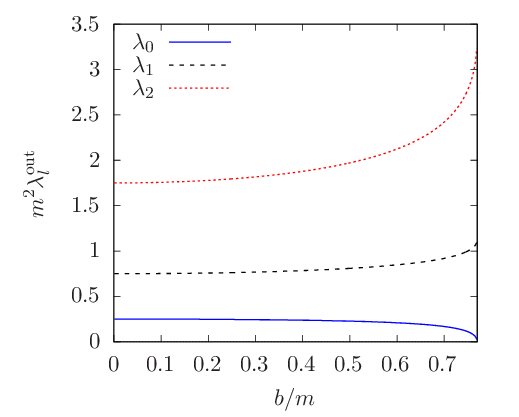}
\caption{Eigenvalues of the spherical MOTS at (a) the inner horizon and (b) outer horizon.}
\label{fig_lambda}
\end{figure}

A MOTS being strictly stable implies the operator admits only positive eigenvalue. In this case, the results of Andersson {\em et al.} established the smooth evolution of the MOTS into a world tube demarcating regions of trapped and untrapped surfaces. The same conclusion on the evolution of the MOTS can be drawn if one weakens the statement on the spectrum to ``the operator admits no vanishing eigenvalue". This is to say that instability of a MOTS does not preclude a smooth evolution of the MOTS. A known example is the case of Reissner-Nordstr\"{o}m black hole with the MOTS of the inner horizon being unstable. This is the case here as well. We see this by noting that for the metric form considered here, these spherical MOTS will have as spectrum for the operator
\begin{align}
\tilde\lambda_{\ell,m}=\frac{\ell(\ell+1)+r_0f'(r_0)}{r_0^2}, \label{StabilityEigenvalue}
\end{align}
with $r_0$ denoting the location of the MOTS, from which, for the Hayward metric, it is verified that the principal eigenvalue for a sphere on the inner horizon is negative. However, it appears that higher eigenvalues are all positive for all $b$ as one would expect. The principal and next two eigenvalues, $\tilde\lambda_1$ and $\tilde\lambda_2$, are plotted against $b$ in Fig.~\ref{fig_lambda} for the inner and outer horizons, respectively.

\begin{figure}[ht!]
 \centering
 \includegraphics[scale=1]{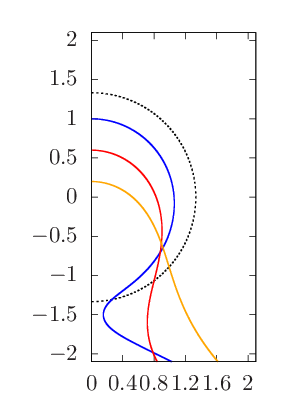}
 \caption{MOTS/MOTOS for the extremal Hayward black hole with $b=\frac{4}{3\sqrt{3}}m$, where its extremal horizon is located at $r=\frac{4}{3}m$, which is a spherical MOTS shown as the black dotted semi-circle. In the interior of the extremal horizon, we have MOTOS shown with initial conditions $r_0=1.0m$, $0.6m$, and $0.2m$, respectively.}
 \label{fig_extremal}
\end{figure}

The behavior of the plots are replicated for higher eigenvalues. As is evident, $\tilde\lambda_1$ and $\tilde\lambda_2$ are positive for all $b$, for both the outer and inner horizons. However, we see that the principal eigenvalue appears to asymptote to zero as one approaches a certain value of $b$. This value is infact that critical value $b=(4/3\sqrt{3})m$ for which the horizons coincide, i.e. the horizon is degenerate. This is a spherical MOTS located at $r=(4/3)m$, and we have generated it from the MOTSodesics equation. We did not find any self-intersecting MOTS/MOTOS to the interior of this degenerate horizon, and we suspect that none can be found. The degenerate horizon and a few of its interior MOTOS are plotted in Fig.~\ref{fig_extremal}. In particular, we suspect that there may be no trapped MOTSodesics in the interior to the degenerate Hayward horizon.

\begin{figure}[ht!]
\centering
    \includegraphics[scale=0.9]{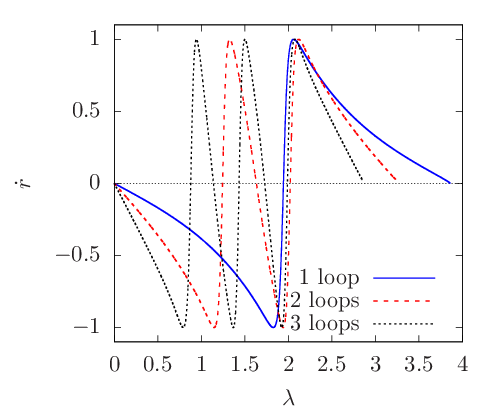}
    \includegraphics[scale=0.9]{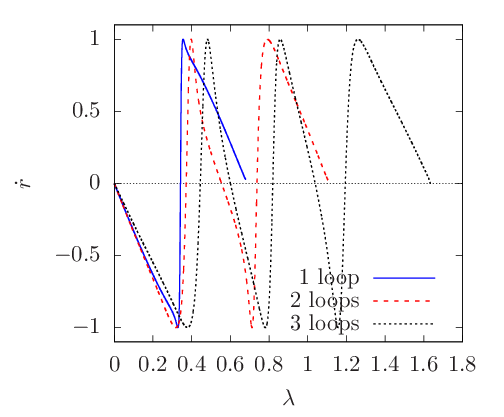}
\caption{$\dot{r}$ vs $\lambda$ for self-intersecting MOTS nearby the outer horizon (left panel) and inner horizon (right panel) respectively.}
\label{fig_rdot}
\end{figure}

The number of negative eigenvalues of the operator for the axi-symmetric MOTS considered here is closely related to the number of loops. In particular, the number of loops places a lower bound on the number of negative eigenvalues (see Theorem 3.4 of \cite{Booth:2023aan} and the follow-up discussions): $\#(\tilde\lambda<0)\geq4n-1$, where $n$ is the number of loops, and the number of loops itself is related to the critical points of $r(\lambda)$. The Fig.~\ref{fig_rdot} shows plots of $\dot{r}$ for three MOTS close to the inner and outer horizons.

In each of the plot we see that for the blue, red, and black $\dot{r}$-curves, there are respectively 1, 3, and 5, critical points, corresponding to 1, 2, and 3 loops, i.e. $\#(\dot{r}=0)=2n-1$. Thus, there are at least, for the respective MOTS, 3, 7, and 11 negative eigenvalues for the operator.

As is seen from the plots of Fig. \ref{fig_data_bstart0.05} and \ref{fig_data_bstart0.3}, and as was alluded to earlier, increasing $b$ corresponds to decreasing pairs of self-intersecting MOTS. The ``size'' of $b$ also carries implications for the number of loops. For small $b$, the frequency of self-intersections of MOTS appears to occur closer to the inner horizon. As the parameter $b$ is increased, we notice that the number of self-intersections decreases rapidly. That is, the parameter $b$, in essence, contains information about the number of negative eigenvalues for the MOTS stability operator. This indicates that the number of self-intersecting MOTS/MOTOS will vanish well inside the outer horizon. It therefore follows that the Theorem 3.4 of \cite{Booth:2023aan}, for counting the least number of negative eigenvalues is not applicable for a relatively large interval of the Hayward parameter space. This was already seen in the case of degeneracy.

\begin{figure}[ht!]
    \includegraphics[scale=0.9]{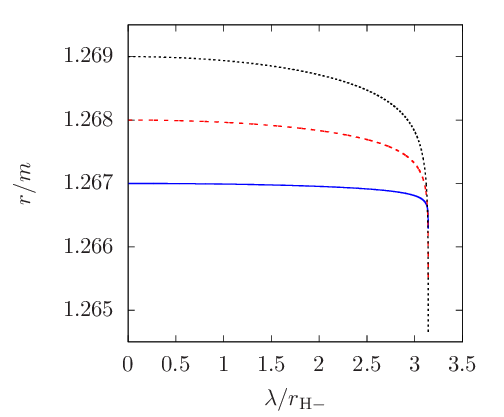}
    \includegraphics[scale=0.9]{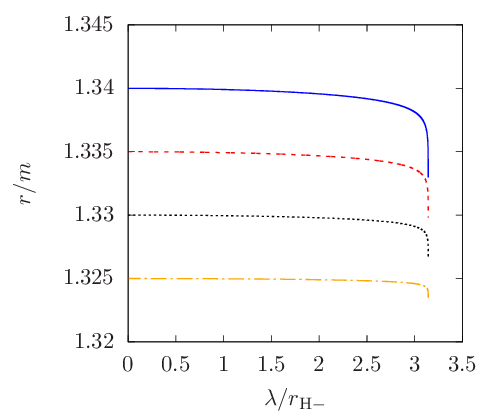}
\caption{Near-spherical and looping MOTS: $b=0.767m$, and near-spherical MOTS: $b=0.7679m$}
\label{fig_NearSph_b0.767_b0.7679}
\end{figure}

The degenerate case informs us that if we decrease $b$ just slightly away from criticality, in which case we still have an inner and an outer horizon, but very close together, the region between them may be populated by a plethora of no self-intersecting MOTS/MOTOS. In fact, the $b$ can be fine-tuned to ensure that only near-spherical (or slightly deformed) MOTS populate the black hole region. In other words, there are values for the parameter $b$ for which the an inner horizon may be continuously deformed into an outer horizon. In Fig.~\ref{fig_NearSph_b0.767_b0.7679} we have graphs of $r$ vs $\lambda$ for near-spherical MOTS and MOTOS in a spacetime with Hayward parameter $b=0.767m$ and $b=0.7679$. For this value the inner and outer horizons are located at $r_{\mathrm{H}-}=1.266609088m$ and $r_{\mathrm{H}+}=1.397901135m$, and $r_{\mathrm{H}-}=1.320864497m$ and $r_{\mathrm{H}+}=1.345724912m$, respectively. The initial conditions from top to bottom are $r_0=1.269m$, $1.268m$, and $1.267m$, and $r_0=1.340m$, $1.335m$, $1.330m,1.325m$, respectively. In the first case, as one gets further away from the inner horizon the MOTS turns and escapes. For the latter all MOTS are nearly-spherial.

To formalize the discussions in the preceding paragraphs, we look for MOTS in the spherical near-horizon regime via a perturbative analysis. As discussed in the previous section, we know that a constant $r=r_d$ and $\theta=\frac{\lambda}{r_d}$ is a solution to the MOTSOdesic equations corresponding to spherical MOTS, provided that $r_d$ coincides with one of the black hole horizons $r_d=r_{\mathrm{H}}$. Though, our interest lies in those cases where for values of a $b$ just below the critical $b$, only near-spherical MOTS (and hence those with no self-intersections) lie in the interior of the black hole.

To do this, we seek MOTS in the neighborhood of these spherical $r_H$ by perturbing about $r_H$. To this end we write
\begin{align}
    r(\lambda)=r_{\mathrm{H}}+\varepsilon r_1(\lambda),\quad \theta(\lambda)=\frac{\lambda}{r_H}+\varepsilon\theta_1(\lambda),
\end{align}
where $\varepsilon$ is a small parameter. Putting these into the MOTSodesic equations \Eqref{EOM_rddot} and \Eqref{EOM_thetaddot}, and expanding to first order in $\varepsilon$ leads to
\begin{subequations}
\begin{align}
    \ddot{r}_1+\frac{1}{r_{\mathrm{H}}}\cot\brac{\frac{\lambda}{r_{\mathrm{H}}}}\dot{r}_1-\frac{1}{r_{\mathrm{H}}}f'(r_{\mathrm{H}})r_1&=0,\label{ddot_r1}\\
    \ddot{\theta}_1+\frac{\dot{r}_1}{r_H^2}&=0,\label{ddot_theta1}
\end{align}
\end{subequations}
where we have used the fact that, near the horizon $f(r_{\mathrm{H}})=0$, we have the expansion for $f$ as
\begin{align*}
    f(r)&\simeq f(r_{\mathrm{H}})+f'(r_{\mathrm{H}})(r-r_{\mathrm{H}})=\varepsilon f'(r_{\mathrm{H}})r_1.
\end{align*}
Notice that the perturbed equations decouple, and so we can work exclusively with Eq.~\Eqref{ddot_r1} to identify the MOTS. Further introducing a change of variables $x=\cos\brac{\lambda/r_H}$, Eq.~\Eqref{ddot_r1} reduces to the Sturm-Liouville-type problem of weight 1.
\begin{align}
    \frac{\dif}{\dif x}\brac{(1-x^2)\frac{\dif r_1}{\dif x}}-r_{\mathrm{H}}f'(r_{\mathrm{H}})r_1&=0.\label{LegendreEq1}
\end{align}
Choosing the parameter $b$, i.e. specifying the Hayward spacetime, fixes $r_H$ (and hence $f'(r_H)$), so that this is an eigenvalue problem, with solutions being the associated eigenfunctions. Consider those choices of the Hayward parameter $b$ for which we can write
\begin{align}
    -r_{\mathrm{H}}f'(r_{\mathrm{H}})=n(n+1),\label{LegendreEigenvaluemain}
\end{align}
where $n$ now denotes the eigenvalues for the operator. It follows that $f'(r_H)$ is strictly negative (see Appendix B). That is, the perturbation is done about the inner horizon. Then, in this case we have that exact solutions to Eq.~\Eqref{LegendreEq1} may be expressed in terms of the hypergeometric function
\begin{align}
_2F_1(-n,n+1;1;x)\qquad \mbox{for}\quad |x|<1,\label{hypergeometric}
\end{align}
i.e.,
\begin{align}
      r_1(\lambda)=c _2F_1\left(-n,n+1;1;\cos\frac{\lambda}{r_H}\right),\label{r1hypergeometric}
\end{align}
for $|\cos\brac{\lambda/r_H}|<1$, where $c$ is an arbitrary constant of order $\sim\mathcal{O}(1)$. The solution $\theta_1(\lambda)$ can then be obtained by successive integrations of Eq.~\Eqref{ddot_theta1}.

One finds that there are only finitely many eigenvalues $n$ for the operator (see Appendix B for an elementary proof of this)
\begin{align}
n\in\{(-2,-1]\cup[0,1)\}.\label{Eigen}
\end{align}
That is, the eigenvalues are not only bounded below but also bounded above. Furthermore, for any two elements $c_1$ and $c_2$ in $(-2,-1)$ and $(0,1)$ respectively, $c_1$ and $-(c_2+1)$ correspond to the same eigenfunction.

The discontinuity in Eq.~\Eqref{Eigen} is a consequence of the allowable horizon values $r_H$, i.e. the restriction on the Hayward parameter $b$ for which the perturbed equation is soluble for the choice \Eqref{LegendreEigenvaluemain}.

\begin{figure}[ht!]
\centering
  \begin{subfigure}[b]{0.49\textwidth}
    \centering
    \includegraphics[scale=0.8]{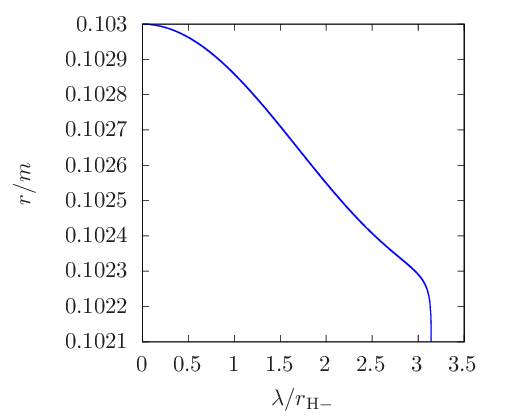}
    \caption{Perturbative solution.}
    \label{fig_PerturbBenchmark_Legendre}
  \end{subfigure}
  \begin{subfigure}[b]{0.49\textwidth}
    \centering
    \includegraphics[scale=0.8]{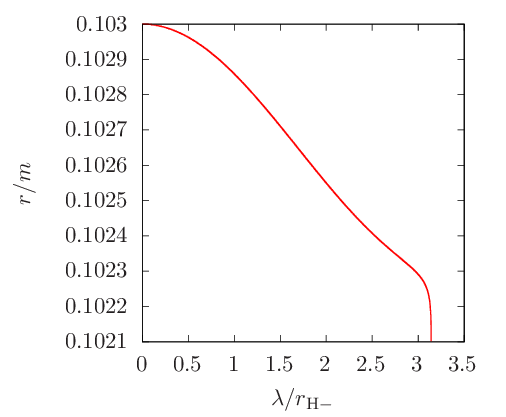}
    \caption{Numerical solution}
    \label{fig_PerturbBenchmark_C}
  \end{subfigure}
  \caption{}
\end{figure}

As a consistency check, one can verify that the perturbative solution obtained here agrees with the numerical solution from integrating \Eqref{ddotr_ddottheta} up to their respective precision/validity. An example for $b=0.1m$ is shown in Fig.~\ref{fig_PerturbBenchmark_Legendre} and Fig.~\ref{fig_PerturbBenchmark_C}. This compares the perturbative solution $r=r_0+\varepsilon c _2F_1\left(-n,n+1;1;\cos\brac{\lambda/r_H}\right)$ and the numerical solution obtained by integrating \Eqref{ddotr_ddottheta} for the case $b=0.1m$. For this value the radius of the inner horizon is located at $r_{\mathrm{H}}=0.1026699997m$. The initial condition is chosen to be $r_0=1.03m$. For the perturbative solution, we have chosen $\varepsilon=10^{-4}$ hence the initial condition fixes $c=3.300003m$.

\section{Conclusion}\label{Conclusion}


In this work we have investigated interior MOTS in Hayward black holes. These are regular and non-vacuum black hole solutions to the Einstein equations, and hence are general enough to warrant an interest since previous works have focused primarily on singular and vacuum solutions. We have numerically generated MOTS and open MOTS for these spacetimes and found a dependence on the number of self-intersecting MOTS and the number of self-intersections of MOTS on the ``size'' of the Hayward parameter $b$. Then, it follows that there is an implicit correlation between the energy density $\rho$ of the spacetime and the self-intersections, although this was not elaborated on. These MOTS occur in pairs as it is with the case of the 4-dimensional Gauss-Bonnet black hole. This pairing therefore appears to be a feature of cases where the coordinate system has a regular ``center''.

The number of self-intersecting MOTS and the number of self-intersections drop very rapidly as one moves away from $b=0$, and as the horizons get closer, i.e as $b$ inches closer to the critical value, it appears to be the case that only near-spherical MOTS exists in the black hole interior. Thus, to look these MOTS is natural to do a perturbation about one of the spherical horizons. (This of course may be done even for small $b$ and generate the near-horizon solutions. We were however motivated by the fact that there is a large space of values for $b$ for which, in principle, all the MOTS may be analytically generated this way.) The perturbed MOTSodesic equations decouple and is reduced to a singular Sturm-Liouville-type problem. Firstly it is found that while the perturbation is adaptable around both the inner and outer horizons, the parameter imposes bound on the horizon radius which in turn imposes that the perturbation is performed only about the inner horizo. Unlike the common practice of writing down analytic solutions to these types of equations in terms of Legendre polynomials, here the solutions are necessarily hypergeometric functions. We checked our analytical results against the numerically generated solutions and they agree to a very fine degree.

This work also builds on the growing literature related to this very interesting subject by extending the analyses into a non-vacuum regime.

An immediate follow-up work is a full investigation into the spectrum of eigenvalues for interior MOTS stability operator. We are quite interested in what constraints the parameter $b$ places on the spectrum. It appears that interesting bifurcation/annihilation processes expected to be apparently generic  to these MOTS may not be present for Hayward black holes. The first indication of this is the non-smooth evolution of the eigenvalues for the horizons. That is, the inner and outer horizons admit no vanishing eigenvalues for all non-critical $b$. This is quite interesting as it deviates from cases that have previously been considered in the literature. It will therefore be of interest to investigate whether there are vanishing eigenvalues for the interior MOTS themselves.


\section*{Acknowledgements}


A.~S would like to thank Ivan Booth for some very useful discussions in understanding aspects of the eigenvalues of the stability operator. A.~S is supported by the Institute of Mathematics, funded through the High-level Talent Research Start-up Project Funding of the Henan Academy of Sciences (Project No.: 251819085). Y.-K.~L is supported by Xiamen University Malaysia Research Fund (Grant no. XMUMRF/2021-C8/IPHY/0001).


\appendix


\section{Deriving the MOTSodesics equations}

The details are provided here for the derivation of the MOTSodesics equations.

\begin{figure}[ht!]
 \centering
 \includegraphics[scale=0.8]{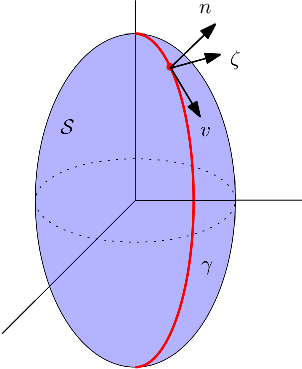}
 \caption{Sketch of a MOTS $\mathcal{S}$ generated as a surface of revolution of a curve $\gamma$ about the vertical axis, where the generator of the revolution is $\zeta$.}
 \label{fig_MOTSodesicsSketch}
\end{figure}

Following \cite{Booth:2021sow}, we assume axi-symmetry and construct $\mathcal{S}$ as a surface of revolution about a symmetry axis in $\Sigma$. To this end, we assume $\Sigma$ has a $U(1)$ isometry generated by a Killing vector $\zeta^a$. (The generating curve $\gamma$ is depicted as the red curve in Fig.~\ref{fig_MOTSodesicsSketch}. The revolution generated by $\zeta^a$ in the figure is a revolution about the vertical axis.)

We take the angular coordinate $\phi$ to be adapted to the integral curves of $\zeta^a$, and the metric on $\Sigma$ shall be written as
\begin{align}
 h_{ab}\dif y^a\dif y^b=\bar{h}_{ij}\dif y^i\dif y^j+R^2\dif\phi^2, \label{axisym_hab}
\end{align}
for $i,j\in\{1,2\}$, where $R$ is some function of $y^i$. We shall assume $\zeta^a$ is normalised so its components are explicitly
\begin{align}
 \zeta^a=\brac{0,0,\frac{1}{R}}.
\end{align}
To construct $\mathcal{S}$, we find a curve $\gamma$ and revolve it about the $\phi$-axis. Therefore we can take $\gamma$ to be lying on the $\phi=0$ plane and is given explicitly in coordinates as
\begin{align}
 \gamma:\mathbb{R}\rightarrow\Sigma,\quad \lambda\mapsto y^a(\lambda)=(y^1(\lambda),y^2(\lambda),0),
\end{align}
where we parametrise the curve by $\lambda\in\mathbb{R}$. The tangent vector to $\gamma$ is
\begin{align}
 v^a=(v^1,v^2,0)=(\dot{y}^1,\dot{y}^2,0),
\end{align}
where over-dots denote derivatives with respect to $\lambda$. Normalize this vector $h_{ab}v^av^b=1$. Now, the vector $\zeta^a$ is normal to $v^a$ and tangent to $\mathcal{S}$ upon revolution of the curve $\gamma$. So the cross product between $\zeta^a$ and $v^a$ will necessarily be normal to $\mathcal{S}$. That is,
\begin{align}
 n_a=\epsilon_{abc}v^b\zeta^c.
\end{align}
The orientation of $n^a$ is chosen such that it points \emph{outward} from $\mathcal{S}$ (and therefore visually consistent with the right-hand rule in Fig.~\ref{fig_MOTSodesicsSketch}), with $\epsilon_{abc}$ being the anti-symmetric Levi--Civita tensor oriented such that $\epsilon_{123}=+\sqrt{\det h}=\sqrt{\det\overline{h}}\;R$. Explicitly, the components of $n_a$ are
\begin{align}
 n_a=\sqrt{\det\bar{h}}\brac{v^2,-v^1,0}.
\end{align}
Since $v^a$ and $\zeta^a$ both have unit norm, it follows from the above that $n^a$ also have unit norm and at a point $p\in\Sigma$, the triplet $\{\zeta^a,v^a,n^a\}$ forms an orthonormal frame which spans $T_p\Sigma$. The projector onto $\mathcal{S}$ then has the decomposition
\begin{align}
 q^{ab}=h^{ab}-n^an^b=v^av^b+\zeta^a\zeta^b,
\end{align}
which is push forward into a tensor on $M$ as
\begin{align}
 q^{\mu\nu}=e^\mu_a e^\nu_b q^{ab}=v^\mu v^\nu+\zeta^\mu\zeta^\nu.\label{q_decomp}
\end{align}

Explicitly one calculates
\begin{align}
z_2=q^{\mu\nu}\nabla_\mu n_\nu&=v^av^b\mathcal{D}_an_b+n^c\partial_c\ln R,\nonumber
&=-\kappa+n^c\partial_c\ln R,
\end{align}
where $\mathcal{D}_an_b=e^\mu_ae^\nu_b\nabla_\mu n_\nu$ is the projected covariant derivative of $n_\mu$ onto $\mathcal{S}$ and $\kappa$ is the magnitude of the acceleration of the curve.

The expression of $z_1$ in terms of the extrinsic curvature follows from the fact that the extrinsic curvature is
\begin{align*}
k_{ab}=\frac{1}{2}\mathcal{L}_uh_{ab},
\end{align*}
with $\mathcal{L}$ being the Lie derivative operator.

In the case of the metric \Eqref{PGstatic1} (we now keep $C$ general), the curve $\gamma$ of interest is parametrized as
\begin{align}
\brac{\tau_0,r,\theta,\phi}=\brac{\tau_0,r(\lambda),\theta(\lambda),\phi}.
\end{align}
The tangent vector to this curve is
\begin{align}
v^a=\brac{\dot{r},\dot{\theta},0},
\end{align}
with $\dot{\ast}=\partial_{\lambda}\ast$. The normalization condition becomes
\begin{align}
v_av^a=C\dot{r}^2+r^2\dot{\theta}^2=1.\label{first_integral}
\end{align}
With the axi-symmetry generator
\begin{align}
\zeta^a=\brac{0,0,\frac{1}{r^2\sin^2\theta}},
\end{align}
the outward normal $n^a$ is
\begin{align}
n^a=\brac{\frac{1}{\sqrt{C}}r\dot{\theta},\sqrt{C}\frac{\dot{r}}{r},0}.
\end{align}
The MOTSodesics equations can now be directly computed:
\begin{subequations} 
\begin{align}
 \ddot{r}&=\frac{-C'\dot{r}^2+2r\dot{\theta}^2}{2C}+\kappa\frac{r\dot{\theta}}{\sqrt{C}},\label{EOM_rddot}\\
 \ddot{\theta}&=-\frac{2}{r}\dot{r}\dot{\theta}-\kappa\frac{\sqrt{C}\dot{r}}{r},\label{EOM_thetaddot}\\
 \kappa&=\frac{(C^2rf'+CC'rf)\dot{r}^2-2(1-Cf)(1+r^2\dot{\theta}^2)}{2r\sqrt{C(1-Cf)}}-\frac{1}{r\sqrt{C}}\brac{\frac{C\cos\theta}{\sin\theta}\dot{r}-r\dot{\theta}}. \label{kappa_eqn}
\end{align}
\end{subequations}
which recovers the equations of Sec.~\ref{Sec:PG} for $C=1$.


\section{The perturberd MOTSodesic equations admit analytic solutions for \texorpdfstring{$n$}{n} in the unit open interval}\label{appendixB}


Here we present elementary arguments to show that the perturbed MOTSodesics equations Eq.~\Eqref{ddot_r1} and Eq.~\Eqref{ddot_theta1} admit analytic solutions for real values $n\in(0,1)$.

Let us set
\begin{align}
    -r_{\mathrm{H}}f'(r_{\mathrm{H}})=n(n+1),\label{LegendreEigenvalue}
\end{align}
with $n$ being real. The horizon equation is given by
\begin{align}
    2mb^2=r_{\mathrm{H}}^2\brac{2m-r_{\mathrm{H}}}. \label{HorizonRelation}
\end{align}
Then, explicitly we have
\begin{align}
   - r_{\mathrm{H}}f'(r_{\mathrm{H}})=-\frac{1}{2m}(3r_{\mathrm{H}}-4m).\label{rHfprimeequation1}
\end{align}
The horizon $r_H$ could be either of the inner or outer horizons, or the extremal one as they are all spherical and hence exact solutions of the ME. However, we note that Eq.~\Eqref{LegendreEigenvalue} is a quadratic in $n$ and a real solution requires the upper bound on $r_H$
\begin{align}
r_H\leq\frac{3}{2}m.\label{bound}
\end{align}
Therefore, the right hand side of Eq.~\Eqref{rHfprimeequation1} is positive, i.e. $f'(r_{\mathrm{H}})<0$. This is to say that $r_H$ can only be one of the inner or extremal horizons. Thus, $n(n+1)\geq0$.
\begin{remark}
From Eq.~\Eqref{rHfprimeequation1}, the bound \Eqref{bound} is sharper
\begin{align*}
r_H\leq\frac{4}{3}m.
\end{align*}
Note that without first establishing $f'(r_{\mathrm{H}})<0$, which requires the choice \Eqref{LegendreEigenvaluemain}, the sharp bound is not necessarily true. For example, without the choice \Eqref{LegendreEigenvaluemain} there could be solutions to the perturbed ME, given appropriate boundary conditions, where $r_H>(4/3)m$ is valid.
\end{remark}

Now, $b=0$ is the Schwarzschild case with the single event horizon, so let us suppose that $b\neq0$. Now, rewrite Eq.~\Eqref{rHfprimeequation1} as
\begin{align}
    -r_{\mathrm{H}}f'(r_{\mathrm{H}})&=n(n+1)=-\frac{b^2}{r_H^2}-\frac{1}{m}(2r_H-3m).\label{rHfprimeequation2}
\end{align}
Since the term involving $b$ is negative, we have
\begin{align}
    n(n+1)<3-\frac{2}{m}r_H<3.\label{rHfprimeequation3}
\end{align}
First consider integer values for $n$. It is immediately clear from Eq.~\Eqref{rHfprimeequation3} that the only allowable positive integer values are $n=\{0,1\}$. It is clear that the $n=0$ mode gives the extremal $r_H$ value for which there are no more two horizons. Of course in this case the perturbed equation is exactly solvable in terms of elementary functions:
\begin{align}
r_1=-\frac{c}{2}\ln\vline\frac{x-1}{x+1}\vline,\label{extremal_perturbed}
\end{align}
for arbitrary constant $c$. Clearly $r_1$ is zero at $x=0$ so that $r=r_H$. That is, these perturbed MOTSodesics from the interior will always cross the extremal horizon as we had earlier suspected.

Furthermore, one checks from Eq.~\Eqref{rHfprimeequation2}, using the horizon equation Eq.~\Eqref{HorizonRelation}, that  $n=1$ gives $r_H=0$, clearly impossible.

In fact, by using the horizon equation in Eq.~\Eqref{rHfprimeequation2} the bound Eq.~\Eqref{rHfprimeequation3} can be sharpened to
\begin{align}
    n(n+1)<2.\label{rHfprimeequation4}
\end{align}
The allowable negative values will then be $n\in(-2,-1]$ (note that $n=-1$ is equivalent to $n=0$ in context here). This then leads to the space of $n$ in Eq.~\Eqref{Eigen}.


\bibliographystyle{mots-Hayward}

\bibliography{mots-Hayward}

\end{document}